

\input harvmac

\Title{BROWN-HET-903}{Matter-Ghost Mixing and the Exact 2d Black Hole Metric}
\centerline{Miao Li}
\bigskip
\centerline{Department of Physics}
\centerline{Brown University}
\centerline{Providence, RI 02912}
\centerline{li@het.brown.edu}
\bigskip
We study a revised version of Witten's 2d black hole, in which the matter
and $(b, c)$ ghosts are mixed. The level of the coset model is still $9/4$.
We show that this model is equivalent to that of Mukhi and Vafa, in which
the level of the coset model is taken as $3$, and the stress tensor is
improved. We argue that the exact metric in such a model is just the
semi-classical one, quite different from the exact metric in Witten's
black hole, being studied by Dijkgraaf, Verlinde and Verlinde. In addition,
there appear ghost-related terms as a part of the background in the world
sheet action.

\Date{4/93}

There has been intensive interest in studying quantum aspects of black holes
recently, partly inspired by Witten's construction of a classically exact
black hole in the two dimensional string theory \ref\witten{E. Witten,
Phys. Rev. D44 (1991) 314.}. The model is provided by a conformal field theory
associated to the coset model $SL(2,R)_k/U(1)$, in case of the Euclidean
black hole, and by the one associated to $SL(2,R)_k/R$, in case of the
minkowskian
black hole. To have a critical central charge $26$, the level $k$ must be
$9/4$. With such a conformal field theory in hand, one hopes to tackle
some problems which may shed light to the role of string theory in quantum
gravity. Though much work has been done along this line \ref\bl{R. Dijkgraaf,
E. Verlinde and H. Verlinde, Nucl. Phys. B371 (1992) 269; M. Bershadsky
and D. Kutasov, Phys. Lett. B266 (1991) 345; E. Martinec and S.L. Shatashvili,
Nucl. Phys. B368 (1992) 338; J. Distler and P. Nelson, Nucl. Phys. B374 (1992)
123.} \ref\mbl{T. Eguchi, H. Kanno and S.-K. Yang, Newton Institute preprint
NI92004; H. Ishikawa and M. Kato, preprint UT-Komaba/92-11.}, we are still
far from any deep understanding of this black hole background. Despite
many interesting problems pertaining to the black hole itself, any further
understanding of it will help us in understanding the 2d string theory,
as the black hole is just one of many possible time-dependent backgrounds
in the 2d string theory.

It was suggested recently by Mukhi and Vafa \ref\mv{S. Mukhi and C. Vafa,
preprint HUTP-93/A002, TIFR/TH/9301.}, that one should not take Witten's
original model as the right black hole background. Instead, one should
consider a modified version in which the matter and the diffeomorphism
$(b,c)$ ghosts are mixed. The recipe is necessary, as argued by these two
authors, in order to have the no-ghost theorem hold in this particular
time-dependent background. The precise construction of this modified model
is the following. One starts with a WZW model $SL(2,R)_3$, with a central
charge $9$, one then improves the stress tensor by adding a term $\partial
J_3$. The resulting theory has a central charge $27$, together with $(b,c)$,
the sum of central charges is $1$. Finally, one takes the $U(1)$ quotient
with the current $J_3+bc$. This theory has zero central charge too. It is
in the last step that the matter and ghost get mixed. We will adopt a
different construction. Our model will be
$${SL(2,R)_{9/4}\otimes [b,c]\over U(1)},$$
the $U(1)$ current is $J'_3+bc$. Here $J'_3$ is a current different from
$J_3$ in the $SL(2,R)$ current algebra, in order to have $J'_3+bc$ be primary.
Note that, here we still start with the
WZW model at level $9/4$, with a central charge $27$.
We shall show however that this model is equivalent to the Mukhi-Vafa model.

One of our main results is that the semi-classical metric, as found in
\witten \ref\tata{G, Mandal, A.M. Sengupta and S. Wadia, Mod. Phys. Lett.
A6 (1991) 1685.}, is the exact metric in the sense that there is no
more $\alpha'$ or $1/k$ corrections. This fact should be contrasted to
the exact metric found in \mbl, where there are $\alpha'$ corrections.
What does this imply to the 2d string theory? This seems to indicate that
the low energy action for the dilaton and metric in fact is the exact
action. A low energy topological model for all topological modes in the 2d
string theory, constructed in \ref\li{M. Li, Brown preprint BROWN-HET-888,
1993.}, seems also to support this result. If one attempts to add $\alpha'$
corrections to that model, there could be no way to preserve topological
invariance.

Now we set off to construct the model. We will keep the level $k$
arbitrary. Let $k'$ denote $k-2$, for $k=9/4$, $k'=1/4$. The current algebra
$(J^\pm, J_3)$ of the $SL(2,R)$ WZW model satisfies
$$\eqalign{J^+(z)J^-(w)=&{k\over (z-w)^2}-{2\over z-w}J_3(w)+\dots,\cr
J_3(z)J^\pm(w)=&\pm{1\over z-w}J^\pm(w)+\dots,\cr
J_3(z)J_3(w)=&-{k/2\over (z-w)^2}+\dots.}$$
To simply life, we shall use a free field realization of the current algebra.
We choose the Wakimoto representation. Three basic free fields are needed.
A pair of bosonic $(\beta, \gamma)$ with spins $(1,0)$, and a scalar $\phi$.
The relevant propagators are $\gamma(z)\beta(w)\sim 1/(z-w)$ and $\phi(z)
\phi(w)\sim -\hbox{ln}(z-w)$. $SL(2,R)_k$ currents are realized as
\eqn\free{\eqalign{J^-(z)=&\beta\gamma^2+\sqrt{2k'}\gamma\partial\phi-
k\partial\gamma,\cr
J^+(z)=&\beta, \quad J_3(z)=\beta\gamma+\sqrt{{k'\over 2}}\partial\phi.}}

To construct the coset, we need to take a quotient with respect to the $U(1)$
group with current $J'_3+bc$. Since $bc$ is not primary, $J'_3$ can not be
$J_3$. We use $J'_3=\beta\gamma-\sqrt{2k'}\partial\phi$, then $J'_3+bc$
becomes primary. We introduce the gauge scalar $X$, and the ghosts
$(\eta,\xi)$ for gauge fixing $U(1)$. These ghosts have spins $(1,0)$.
Associated to current $J'_3+bc$, the anomaly free current is $J'_3+bc
-i\sqrt{2k'}\partial X$. The radius of $X$ is $\sqrt{2k'}$, and becomes
self-dual when $k'=1/4$. The $U(1)$ BRST charge is then
$$Q_{U(1)}=\oint\xi(J_3+bc-i\sqrt{2k'}\partial X).$$
Throughout this letter, $\oint$ stands for $1/(2\pi i)\oint dz$. This BRST
charge is nilpotent.

Now, the total stress tensor including the $(b,c)$ ghosts is
$$T(z)=-\beta\partial\gamma - \eta\partial \xi -{1\over 2}(\partial X)^2
-{1\over 2}(\partial\phi)^2+{1\over\sqrt{2k'}}
\partial^2\phi-2b\partial c+c\partial b $$
The central charge of $(\beta,\gamma)$ and $(\eta, \xi)$ sums to zero, it
is reasonable to expect that one can eliminate these fields effectively.
We adopt a similar transformation on the stress tensor as found by Eguchi
et al. in \mbl. Consider the BRST anti-commutator
$$
\{Q_{U(1)}, \eta\partial\hbox{ln}\gamma\}=\beta\partial \gamma+\eta\partial
\xi +(bc-\sqrt{2k'}\left(\partial\phi+i\partial X)\right)\partial
\hbox{ln}\gamma+{1\over 2}(\partial\hbox{ln}\gamma)^2-{1\over 2}\partial^2
\hbox{ln}\gamma.$$
Note that $\partial\hbox{ln}\gamma$ is a conformal operator. The above formula
suggests us to add it to the stress tensor to cancel out the kinetic part
for $(\beta, \gamma)$ and $(\eta, \xi)$. We arrive at a new stress tensor
$T'$, equivalent to $T$:
\eqn\new{\eqalign{T'=&T+\{Q_{U(1)}, \eta\partial\hbox{ln}\gamma\}\cr
=&-{1\over2}(\partial X')^2-{1\over 2}(\partial\phi')^2+{1\over\sqrt{2k'}}
\partial^2\phi'-2b'\partial c'+c'\partial b'+\partial^2\hbox{ln}\gamma,}}
where
\eqn\trans{\eqalign{X'=&X+i\sqrt{2k'}\hbox{ln}\gamma,\quad \phi'=\phi
+\sqrt{2k'}\hbox{ln}\gamma,\cr
b'=&b\gamma,\quad c'=c\gamma^{-1}.}}
We note that with the new fields defined above, the new stress tensor in \new\
is almost the same as the one for the c=1 Liouville theory, when $k'=1/4$.
The only difference is the last term associated to $\gamma$. Since there is
no kinetic term for $\gamma$ in the new stress tensor, $\gamma$ is no longer
dynamic. It is then consistent to take $\gamma$ being constant, if one works
with new fields and new stress tensor. Thus, the total derivative term
associated to $\gamma$ can be dropped out. This phenomenon is slightly
different
from that in the work of Eguchi et al, where no such a total derivative term
is left after do a similar BRST transformation. It is easy to check that all
new fields are neutral with respect to $Q_{U(1)}$, so operators constructed
from them are automatically BRST invariant. Note that, if we set $\gamma$
be constant, then one can replace $c$ in the diffeomorphism BRST operator
by $c'$.

The stress tensor is not the whole story of the coset model. To calculate
correlation functions, a screening operator is needed. Or equivalently, one
adds a (1,1) operator constructed from the screening operator to the world
sheet lagrangian. This operator in the present model is
\eqn\scre{(\beta\gamma^3-3\gamma\partial\gamma)(\overline{\beta}\overline{
\gamma}^3-3\overline{\gamma}\overline{\partial}\overline{\gamma})
e^{\sqrt{2/k'}\phi}.}
This operator is invariant with respect to $Q_{U(1)}$ (in the
ordinary model where there is no matter-ghost mixing, the screening operator
is $\beta\overline{\beta}\hbox{exp}(\sqrt{2/k'}\phi)$). To find the
corresponding operator depending only on new fields, we use
$$\{Q_{U(1)}, \eta\gamma^2e^{\sqrt{2/k'}\phi}\}=(\beta\gamma^3-3\gamma\partial
\gamma) e^{\sqrt{2/k'}
\phi}+\sqrt{2k'}\left(-\partial (\phi'+iX')+{1\over\sqrt{2k'}}b'c'\right)
e^{\sqrt{2/k'}\phi'}.$$
So the first operator
on the r.h.s. of the above equation is equivalent to the second operator on
the r.h.s., up to a BRST anti-commutator. One can go a step further to show
that the operator
\eqn\screen{V=\left(\partial (\phi'+iX')-{1\over \sqrt{2k'}}b'c'\right)\left(
\overline{\partial}(\phi'+iX')-{1\over\sqrt{ 2k'}}\bar{b}'\bar{c}'\right)
e^{\sqrt{2/k'}\phi'}}
is equivalent to \scre, up to a BRST anti-commutator in terms of the total BRST
charge $Q_{U(1)}+\overline{Q}_{U(1)}$. $V$ is constructed only from new
fields, therefore should be used when one works with new stress tensor and new
fields. It is remarkable that this operator contains ghosts explicitly,
this is how the matter and ghost get mixed in terms of new fields. We have
cheated a little in the above discussion. Both operators in \scre\ and in
\screen\ are not primary operators. It is easy to see that $V$ is not primary,
since $b'c'$ is not. We need certain prescription for calculating correlation
functions in order to circumvent this problem. There is one, that is, to
use the twisted N=2 CFT, as the present model can be thought of as a twisted
N=2 CFT \mv.

We now discuss the world sheet action in terms of new fields. The purely
kinetic part is
$$S_0={1\over 2\pi}\int dz^2(\partial X'\overline{\partial} X'+\partial\phi'
\overline{\partial}\phi'+\sqrt{{2\over k'}}R\phi'),$$
where we omitted the kinetic term for ghosts $(b', c')$. The dilaton field
is $\Phi=\sqrt{1/2k'}\phi$. We add a term $\mu/(2\pi)V$ to the lagrangian, the
action for bosonic fields  now reads
\eqn\act{S={1\over 2\pi}\int dz^2(g_{X'X'}\partial X'\overline{\partial}
X'+g_{\phi'\phi'}\partial\phi'\overline{\partial}\phi'+2g_{X'\phi'}\partial
X'\overline{\partial}\phi'+\sqrt{{2\over k'}}R\phi'),}
with metric components
$$\eqalign{g_{X'X'}=&1-\mu e^{\sqrt{2/k'}\phi'},\quad g_{\phi'\phi'}
=1+\mu e^{\sqrt{2/k'}\phi'},\cr
g_{X'\phi'}=&i\mu e^{\sqrt{2/k'}\phi'}}.$$
Since the screening operator is believed to be exactly marginal, the metric
we get should be taken as exact. The above metric can be diagonalized. If we
define new coordinates via $\hbox{tanh}^2r=1-\mu\hbox{exp}(\sqrt{2/k'}\phi'
)$, and $\theta= X'-i\sqrt{2k'}\hbox{ln}(\hbox{tanh}r)$, the diagonalized
metric is
\eqn\bl{g_{rr}=2k', \quad g_{\theta\theta}=\hbox{tanh}^2r,}
the standard euclidean black hole metric, found as a solution to the low
energy effective action. The dilaton field in term of new coordinates is
$2\Phi=-\hbox{ln}(\mu\hbox{cosh}^2 r)$, also the standard one. The radius of
$\theta$ is $\sqrt{2k'}$, the same as that of $X'$. If $X'$ were real, $
\theta$ would be complex. To get rid of this embarrassing situation, one
can Wick-rotate $X'$ and $\theta$, thus gets the minkowskian black hole.
Why is the exact metric the same as the solution to the low energy effective
action? The answer is that there is another part in the world sheet action,
related to $(b', c')$ ghosts. This part is written, in terms of new
coordinates, as
$$\eqalign{{1\over 2\pi}\int dz^2&{1\over\sqrt{2k'}}{1\over\hbox{cosh}^2r}
[(-i\partial \theta+\sqrt{2k'}\hbox{coth}r\partial r)\bar{b}'\bar{c}'
+(-i\overline{\partial}\theta+\sqrt{2k'}\hbox{coth}r\overline{\partial}
r)b'c'\cr
&-{1\over\sqrt{2k'}}b'\bar{b}'c'\bar{c}'].}$$
Indeed this part together with that for bosonic fields gives us almost
a N=2 system, except that the spins of $(b',c')$ are not $(1/2, 1/2)$
but $(2,-1)$. The model can be treated as a twisted N=2 system, with
fermionic fields $(b', c')$. The hidden N=2 symmetry is the
reason why the beta functions for the dilaton and metric calculated at
semi-classical level are exact. The above term involving ghosts can be
compared to that discussed in \ref\nojiri{S. Nojiri, Phys. Lett. B274
(1992) 41.}, where a supersymmetric gauged WZW model based on group
$SL(2,R)$ is considered. If one replaces $k$ by $k'=k-2$ in those
semiclassical formulas in \nojiri, one gets the same metric and the
ghost-related terms. So the only renormalization effect is the shift of
the level.

In the Witten's version of the 2d euclidean black hole, it was shown
in the first paper in
\mbl\ that the screening operator corresponds to the wrongly dressed
operator $W^{-}_{1,0}$ in the c=1 model. What does our screening operator
in the revised model correspond to? In the first sight it seems impossible
to find any corresponding operator in the c=1 model, since our screening
operator contains ghosts. The right way is to consider the conformal dimension
zero, BRST invariant operator $c'\bar{c}'V$. It is the factor $c'\bar{c}'$
that annihilates the ghost terms in $V$. Still, this operator is different
from that in the Witten's model. To see that the screening operator is
indeed equivalent to $c'\bar{c}'W^-_{1,0}$, we note that $W^-_{1,0}\sim
\partial X\overline{\partial}X\hbox{exp}(\sqrt{2/k'}\phi')$. In addition
to this term, there are terms
in $c'\bar{c'}V$ proportional to $\partial\phi'\hbox{exp}(\sqrt{2/k'}\phi')$
or its anti-holomorphic counterpart. Let $Q'$ be the diffeomorphism BRST
operator constructed from $(b', c')$ and the new stress tensor \new, then one
can show that
$$c'\partial \phi' e^{\sqrt{2/k'}\phi'}=[Q', \sqrt{{k'\over 2}} e^{
\sqrt{2/k'}\phi'}].$$
Thus, the operator $c'\bar{c}'V$ is equivalent to $c'\bar{c}'W^-_{1,0}$.
It is remarkable that both screening operators in the matter-ghost mixing
model and the non-mixing model correspond to the same BRST invariant
operator in the c=1 model, while they are markedly different as $(1,1)$
operators.

As we showed before, when $\gamma$ is effectively dropped out after doing
transformation in \new, one can use $Q'$ to calculate the BRST cohomology.
This BRST cohomology is just the same as in the c=1 Liouville model
\ref\brst{B. Lian and G. Zuckerman, Phys. Lett. B266 (1991) 21;
P. Bouwknegt, J. McCarthy and K. Pilch, Comm. Math. Phys. 145 (1992)
541.}. One can use the old fields $\beta, \gamma, \phi$ and $X$, and $b, c$
to express those operators. However, when one considers $(1,1)$ operators
which can be used to perturb the string background, one should be more
careful. The simplest example is just $V$ in \screen. This operator
provides a time-dependent background, and simultaneously matter-ghost mixing.
It is the latter that is missing in the corresponding BRST invariant operator,
as we showed in the last paragraph. It is generally true that any $(1,1)$
discrete operator gives rise to a time-dependent background, therefore
one should carefully find out the ghost-dependent piece.

The tachyon vertex operator $\hbox{exp}(ip_XX'+p_\phi\phi')$ is
$$\gamma^{\sqrt{2k'}(p_\phi-p_X)}e^{ip_XX+ip_\phi\phi}$$
in terms of $SL(2,R)$ free fields. This is also different from the one in the
ordinary model without matter-ghost mixing \mbl. Given that both the tachyon
vertex operator and the screening operator $(\beta\gamma^3-3\gamma\partial
\gamma)(\overline{\beta}\overline{\gamma}^3-3\overline{\gamma}\overline{
\partial}\overline{\gamma})\hbox{exp}(\sqrt{2/k'}\phi)$
are different from those in
the ordinary model, we expect that tachyon amplitudes will be different
in these two models, even at tree level. It would be extremely interesting
to calculate just two point amplitudes in both models and compare them.

Next we show that the model proposed by Mukhi and Vafa in \mv\ is equivalent
to ours. Their starting point is a WZW model $SL(2,R)_3$, with the improved
stress tensor $T+\partial J_3$. The central charge of this model is again
$27$. We shall replace the level $3$ by $k+2$, for a general $k$. Once again
we use fields $(\beta,\gamma)$ and $\phi$ to bosonize the current algebra.
We have to replace $k'$ by $k$ and $k$ by $k+2$ in formulas in \free. With the
improvement, the stress tensor for this WZW model is
\eqn\stress{T(z)=\gamma\partial\beta-{1\over 2}(\partial\phi)^2+{1\over
\sqrt{2}}(\sqrt{k}+1/\sqrt{k})\partial^2\phi.}
Now the spins of $(\beta,\gamma)$ after improvement are longer $(1,0)$ but
$(0,1)$, so the roles of $\beta$ and $\gamma$ are exchanged. This implies
particularly that current $J^+$ is of conformal dimension $0$. Define
$1/\sqrt{k'}=\sqrt{k}+1/\sqrt{k}$, then the stress tensor in \stress\ is\
the same as that of $SL(2,R)$ at level $k'+2$. Taking $k=1$, the value
proposed in \mv, $k'=1/4$, the value in our model studied before. This is
the main reason why the two models are the same. The $U(1)$ current to be
modded out is $-J_3+bc$. Although both $J_3$ and $bc$ are not primary with
the improved stress tensor, $-J_3+bc$ is. Now the $U(1)$ BRST charge is
$$Q_{U(1)}=\oint \xi (-J_3+bc-i\sqrt{{k\over 2}}\partial X).$$
Again taking $k=1$, we find that this BRST charge is the same as the one we
defined
before when $k'=1/4$, provided we use $\beta\rightarrow \gamma$ and $\gamma
\rightarrow -\beta$. The statement that with the improved stress tensor, one
puts constraint $J^+=\beta=\hbox{const}$ is equivalent to what we had to do
with $\gamma$ in our model, in order to get rid the total derivative term
in the stress tensor $T'$ in \new.

To conclude, we have shown that in the revised version of Witten's black hole,
the semi-classical metric and dilaton become exact, due to the appearance of
ghost-related terms in the world sheet action.
The necessity of the mixing of the matter and ghosts can be most easily seen
by examining the one-loop partition function, where appropriate cancellation
has to occur in order to have the right amount of degrees of freedom
\ref\cumrun{C. Vafa, talk given at SUSY'93.}. If the lesson we learned from
this example is correct, we have to be careful when we consider any
time-dependent background, obtained by adding marginal perturbation to the
action. The marginal operator should contain ghost-related terms in an
appropriate way. The guide for introducing these terms may be some hidden
N=2 supersymmetry, as advocated by the authors of \mv. Indeed the hidden N=2
symmetry is the reason why the semi-classical solution is exact in the model
discussed in this letter. Consider a N=2 non-linear sigma model with a
dilaton field $\Phi$, the Ricci tensor associated to the rescaled metric
$\hbox{exp}(-2\Phi)g$ should vanish. This condition is exact for the N=2
non-linear sigma model. When the spacetime is two dimensional, this
implies that the rescaled metric is flat, a consequence of the semi-classical
equation of motion.

\noindent {\bf Acknowledgments}

It is a pleasure to thank A. Jevicki for discussions concerning many issues
of black holes and string theory. I am grateful to C. Vafa for
emphasizing to me the necessity of matter-ghost mixing. This work was
supported by DOE contract DE-FG02-91ER40688-Task A, and part of this work
was done during the author's visit at ITP, Santa Barbara.

\listrefs\end